\documentclass{PoS}

\def\ga{\mathrel{\raise.3ex\hbox{$>$\kern-.75em\lower1ex\hbox{$\sim$}}}}
\def\la{\mathrel{\raise.3ex\hbox{$<$\kern-.75em\lower1ex\hbox{$\sim$}}}}

\newcommand{\lam}{\lambda}
\def\lsim{\mathrel{\rlap{\lower4pt\hbox{\hskip1pt$\sim$}}
    \raise1pt\hbox{$<$}}}                
\def\gsim{\mathrel{\rlap{\lower4pt\hbox{\hskip1pt$\sim$}}
    \raise1pt\hbox{$>$}}}                

\title{Charged Higgs Boson Benchmarks in the CP-conserving 2HDM}

\ShortTitle{Charged Higgs in 2HDM}

\author{Renato Guedes\\
        Centro de F\'\i sica Te\' orica e Computacional, Faculdade de Ci\^encias, Universidade de Lisboa, Av. Prof. Gama Pinto 2, 1649-003 Lisboa, Portugal. \\
        E-mail: \email{renato@cii.fc.ul.pt}}
        
\author{Shinya Kanemura\\
        Department of Physics, The University of Toyama, 3190 Gofuku, Toyama 930-8555, Japan.\\
        E-mail: \email{kanemu@sci.u-toyama.ac.jp}}       
        
\author{Stefano Moretti \\
        NExT Institute and School of Physics and
Astronomy, University of Southampton Highfield, Southampton SO17
1BJ, UK.\\
        E-mail: \email{stefano@phys.soton.ac.uk}}       

\author{\speaker{Rui Santos}
\\
        Instituto Superior de Engenharia de Lisboa, Rua Conselheiro Em\'\i dio Navarro 1, 1959-007 Lisboa, Portugal \textit{and} \\
        Centro de F\'\i sica Te\' orica e Computacional, Faculdade de Ci\^encias, Universidade de Lisboa, Av. Prof. Gama Pinto 2, 1649-003 Lisboa, Portugal. \\
        E-mail: \email{rsantos@cii.fc.ul.pt}}

\author{Kei Yagyu\\
        Department of Physics, The University of Toyama, 3190 Gofuku, Toyama 930-8555, Japan.\\
        E-mail: \email{keiyagyu@jodo.sci.u-toyama.ac.jp}}       

\abstract{We present a discussion on charged Higgs boson searches at the Large Hadron Collider in CP-conserving two-Higgs Doublet Models.}

\FullConference{Third International Workshop on Prospects for Charged Higgs Discovery at Colliders - CHARGED2010,\\
		September 27-30, 2010\\
		Uppsala Sweden}

\begin{document}

\section{Introduction}

\noindent
This is a a contribution to the Charged Higgs Benchmark working group: general 2HDM. We present a general discussion of the the different Yukawa types of a CP-conserving model with no Flavour Changing Neutral Currents (FCNCs) at tree-level as well as their experimental and theoretical constraints. We then proceed to define the production and decay processes to be studied. We give an overview on how the different processes can probe each version of the model and also if a distinction between the different model versions and also between each model version and the MSSM can be expected for some regions of the parameter space. 

\section{The two-Higgs doublet model}

\noindent
We start with a brief review of the versions of the 2HDM used in this work. We have chosen a potential with a minimum that does not break CP-invariance nor electric charge and that was shown to be stable at tree level~\cite{vacstab}. Under these constraints, the most general renormalizable potential which is invariant under $SU(2) \otimes U(1)$ can be written as
\begin{eqnarray}
V(\Phi_1,\Phi_2) &=& m^2_1 \Phi^{\dagger}_1\Phi_1+m^2_2
\Phi^{\dagger}_2\Phi_2 - (m^2_{12} \Phi^{\dagger}_1\Phi_2+{\rm
h.c}) +\frac{1}{2} \lam_1 (\Phi^{\dagger}_1\Phi_1)^2 +\frac{1}{2}
\lam_2 (\Phi^{\dagger}_2\Phi_2)^2\nonumber \\ &+& \lam_3
(\Phi^{\dagger}_1\Phi_1)(\Phi^{\dagger}_2\Phi_2) + \lam_4
(\Phi^{\dagger}_1\Phi_2)(\Phi^{\dagger}_2\Phi_1) + \frac{1}{2}
\lam_5[(\Phi^{\dagger}_1\Phi_2)^2+{\rm h.c.}] ~, \label{higgspot}
\end{eqnarray}
where $\Phi_i$, $i=1,2$ are complex $SU(2)$ doublets with four degrees of freedom each and all $m_{i}^2$, $\lambda_i$ and $m_{12}^2$ are real. Once the $SU(2)$ symmetry is broken,  we end up with two CP-even Higgs states $h$ and $H$, one CP-odd state, $A$, two charged Higgs bosons, $H^{\pm}$ and three Goldstone bosons. This potential has seven independent parameters which we choose to be the four masses $m_{h}$, $m_{H}$, $m_{A}$, $m_{H^\pm}$, $\tan\beta=v_2/v_1$, $\alpha$ and $M^2$. The angle $\beta$ is the rotation angle from the group eigenstates to the mass eigenstates in the CP-odd and charged Higgs sector. The angle $\alpha$ is the corresponding rotation angle for the CP-even Higgs sector. The parameter $M^2$ is defined as $M^2=m_{12}^2/(\sin \beta \cos \beta)$ and is a measure of how the discrete symmetry is broken. The discrete symmetry imposed to the potential, when extended to the Yukawa Lagrangian, guarantees that FCNCs are not present as fermions of a given electric charge couple to no more than one Higgs doublet~\cite{Glashow}. There are a total of four possible combinations~\cite{barger} and therefore four variations of the model. We define as Type I the model where only the doublet $\phi_2$ couples to all fermions; Type II is the model where $\phi_2$ couples to up-type quarks and $\phi_1$ couples to down-type quarks and leptons; a Type Y~\cite{KY} or III model is built such that $\phi_2$ couples to up-type quarks and to leptons and $\phi_1$ couples to down-type quarks and finally in a Type X~\cite{KY} or IV model, $\phi_2$ couples to all quarks and $\phi_1$ couples to all leptons.

\begin{figure}[h]
\centering
\includegraphics[height=2.87in]{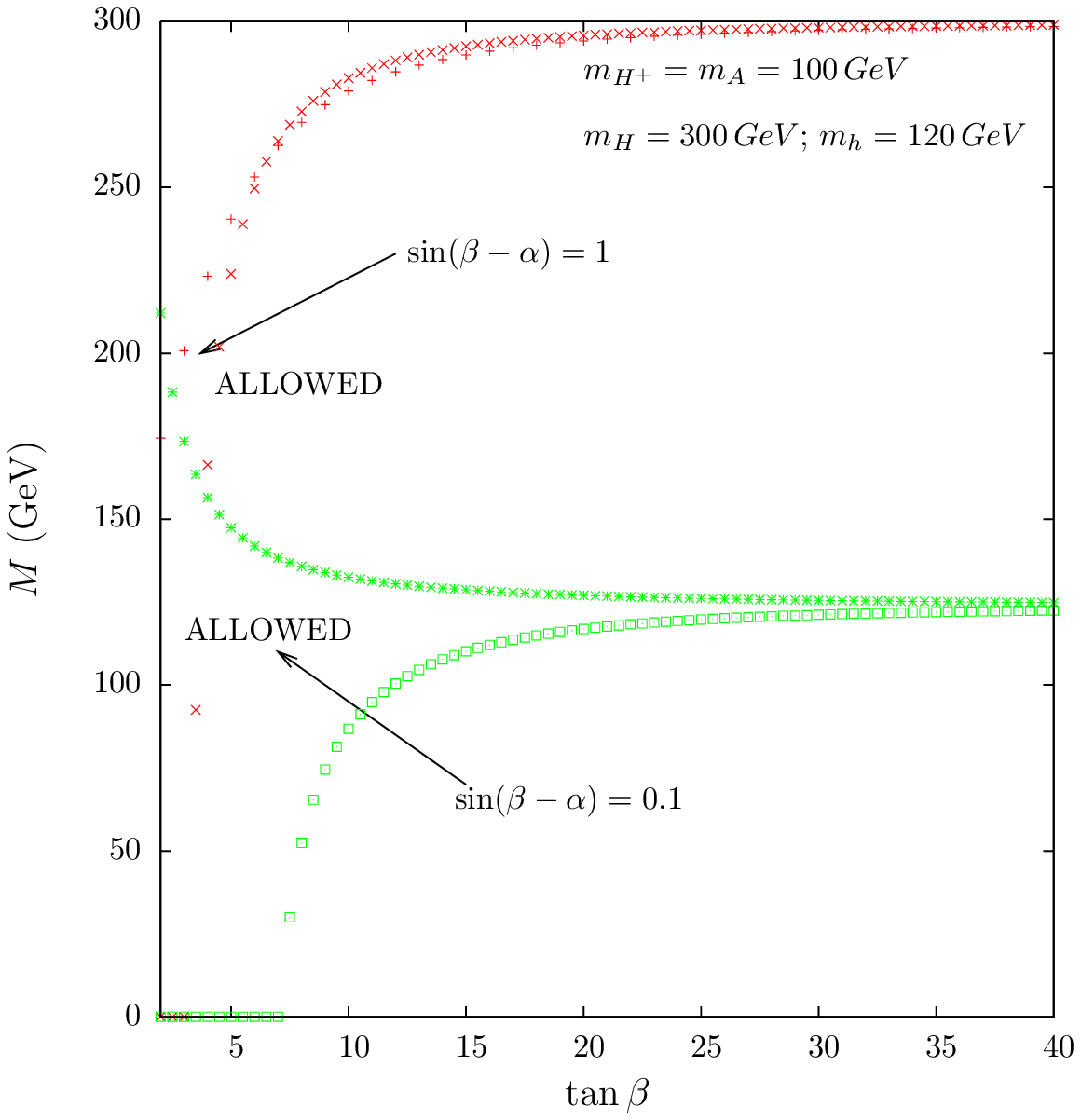}
\includegraphics[height=2.9in]{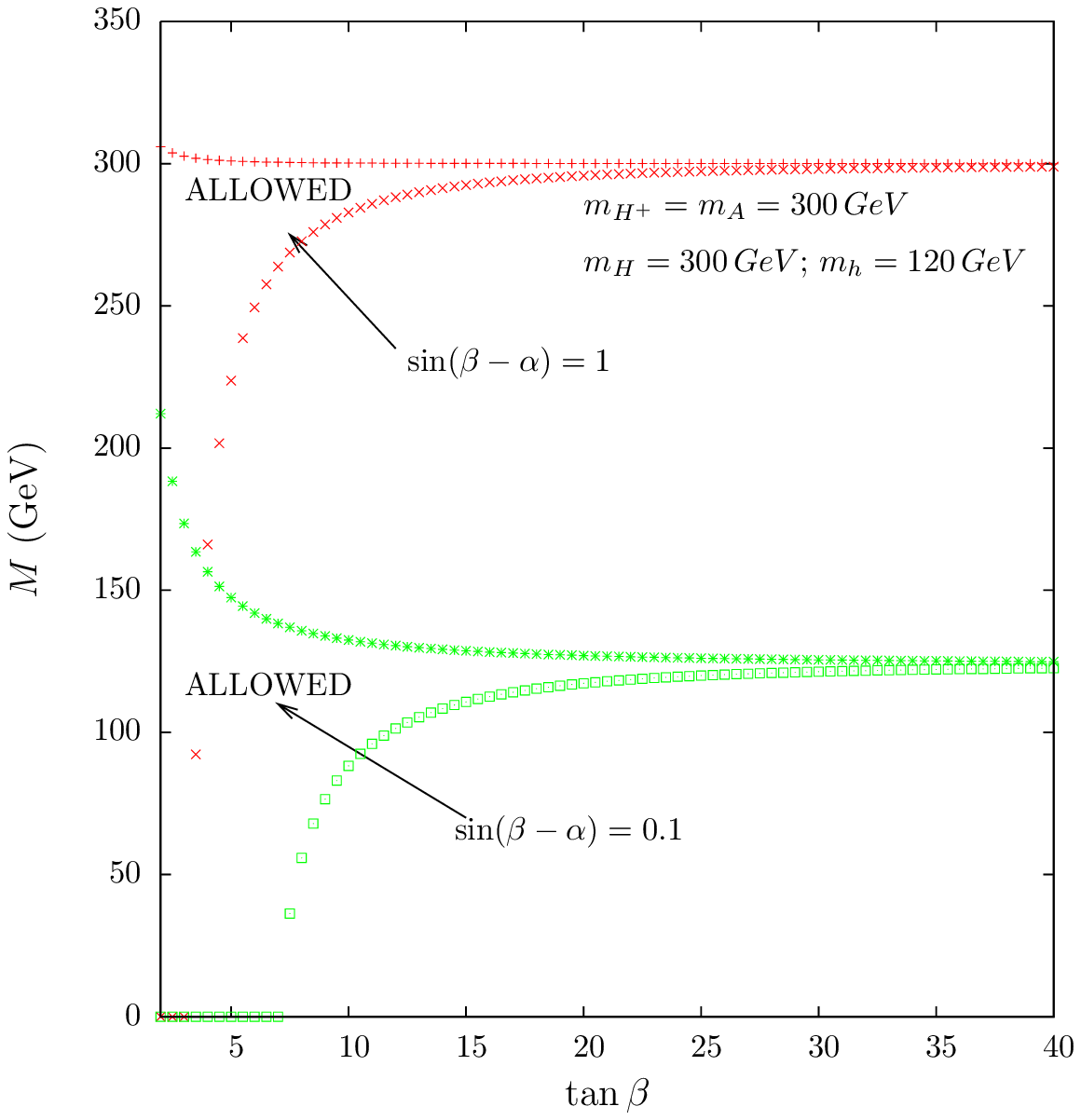}
\caption {Allowed regions of the parameter space when vacuum stability and perturbative unitarity are imposed on the model.}
\label{fig:allowed}
\vskip -0.6cm
\end{figure}

\noindent
We will now very briefly discuss the main experimental and theoretical constraints which affect the parameter space of these 2HDM types. New contributions to the $\rho$ parameter stemming from Higgs states \cite{Rhoparam} have to comply with the current limits from precision measurements \cite{pdg}: $ |\delta\rho| \la 10^{-3}$; values of $\tan \beta$ smaller than $\approx 1$ together with a charged Higgs with a mass below 100 GeV are disallowed both by the constraints coming from $R_b$ and from $B_q \bar{B_q}$ mixing~\cite{osland} for all Yukawa versions of the model. It has been shown in~\cite{bsgamma} that data from $B\to X_s \gamma$ impose a lower limit of $m_{H^\pm} \ga 300$ GeV in models where the quarks have Type II or Type Y Yukawa couplings. In models Type I and Type X charged Higgs bosons as light as 100 GeV are still allowed. The LEP experiments have set a lower limit on the mass of the charged Higgs boson 79.3 GeV  at 95\% C.L., assuming only $BR(H^+ \to \tau^+ \nu) + BR(H^+ \to c \bar s)=1$~\cite{LEP}. The limit becomes stronger if $BR(H^+ \to \tau^+ \nu)  \approx 1$ (see \cite{Logan:2009uf} for a discussion and a review on bounds for model X). These bounds led us to choose two benchmarks for the masses  - 100 GeV for models I and X and 300 GeV for the four Yukawa versions of the model - the ones that maximise the cross section while complying with all bounds. We will vary $\tan \beta$ from 1 to 40. The allowed values of the remaining masses, of $M$, $\sin \alpha$ will be mainly contrained by tree-level vacuum stability~\cite{vac1} and tree-level unitarity~\cite{unitarity} of the potential. In Fig.~\ref{fig:allowed} we show how vacuum stability and perturbative unitarity constrain the parameter space of the model. In the left panel we plot $M$ as a function of $\tan \beta$ for $m_{H^\pm}=100$ GeV and two values of $\sin (\beta - \alpha)$. In the right panel we now take $m_{H^\pm}=300$ GeV. The values of the remaining parameters are shown in the plots. The most striking feature is that $\tan \beta$ can be constrained to be very small, independently of the value of $M$. In any case as $\tan \beta$ grows the allowed values of $M$ shrink to a tiny region that depends mainly on $m_H$ and $\sin (\beta - \alpha)$. 

\section{Charged Higgs production and decay}

\noindent
This section describes all relevant production and decay channels of charged Higgs bosons in all 2HDM types at the LHC.  We will start by listing the most relevant production modes. We can have single $H^\pm$ production \cite{btH,bQH,cs,WH,WHb}:
\begin{eqnarray}\label{Eq:single-production}
gg,q\bar q&\to b \bar t H^+ \textsl{ and } \, \bar b t H^-, \label{Eq:btH} \\
bQ        &\to bQ'H^+ \textsl{ and }  \, bQ'H^-, \label{Eq:bQH}\\
cs        &\to H^{\pm}  (+jet), \label{Eq:qQH}\\
gg,b\bar b&\to W^-H^+ \textsl{ and } \,  W^+H^-, \label{Eq:WH}
\end{eqnarray}
or pair production \cite{HpHm,DY,VBF}
\begin{eqnarray}\label{Eq:pair-production}
gg,b\bar b&\to H^+H^-, \label{Eq:HH} \\
q\bar q   &\to H^+H^-, \label{Eq:DY} \\
qQ        &\to q'Q' H^+H^-, \label{Eq:VBF}
\end{eqnarray}
where $q,q^\prime,Q,Q^\prime$ represent (anti)quarks (other than $b$'s and $t$'s). We do not consider associated production of a charged Higgs with another scalar. As stated before we will concentrate on charged Higgs boson masses of 100 GeV for models I and X  and 300 GeV for all models. We use as benchmarks the values of $\tan \beta =1, \, 3, \, 7$ and $30$. The production cross sections are the same for models I and X and for models II and Y due to the coupling of the Higgs fields to the quarks. The remaining values for the benchmarks are defined as to cover the main different signatures that arise in each model. 
\begin{table}[ht]
\begin{center}
\begin{tabular}{c c c c c c c c c c c c c c c c c c c c c c c c } \hline \hline
$\tan \beta$ && $\sigma$ (pb)   &&  I $(cs)$ (pb) &&  I $( \tau \nu)$ (pb)   && X $(cs)$ (pb)  &&  X $( \tau \nu)$ (pb)   \\
\hline
1        &&  522     &&    162      &&  360      &&   162    &&  360    \\ \hline
3        &&  82       &&    26     &&  56     &&   0    && 82     \\ \hline
7        &&  15.6    &&    4.9     &&  10.7     &&   0    && 15.6      \\ \hline
30      &&  0.86    &&    0.27   &&  0.59     &&   0  && 0.86      \\ \hline \hline
\end{tabular}
\caption{Cross sections for the different models and for the main charged Higgs decays for $m_{H^\pm} = m_A = 100 $ GeV and four values of $\tan \beta$. Remaining parameters are $m_H = M = 300$ GeV, $m_h = 120$ GeV and $\sin (\beta - \alpha) = 1$.}
\end{center}
\vskip -0.6cm
\end{table}
\label{tab:tab100}

\noindent
The first process \ref{Eq:btH} on our list depends only on $\tan \beta$ and on the charged Higgs mass. 
\begin{table}[h!]
\begin{center}
\begin{tabular}{c c c c c c c c c c c c c c c c c c c c c c c c } \hline \hline
$\tan \beta$ && $\sigma_{I,X}$ (fb) && $\sigma_{II,Y}$ (fb)   &&  I (fb) &&  II  (fb)   && X (fb)  &&  Y  (fb)   \\
\hline
1        &&  2189  &&  2178    &&    2189 ($tb$)     &&  2178 ($tb$)     &&   2189 ($tb$)   &&  2178 ($tb$) \\ \hline
3        &&  244    &&  247      &&    244 ($tb$)       &&  247 ($tb$)       &&   244 ($tb$)     && 247 ($tb$)       \\ \hline
7        &&  44      &&  72        &&    44 ($tb$)         &&  64 ($tb$)         &&   39 ($tb$)       && 72 ($tb$)      \\ \hline
30      &&  2.4     &&  562      &&    2.4 ($tb$)        &&  433 ($tb$)      &&   2.4 ($\tau \nu$)      && 562 ($tb$)      \\ \hline \hline
\end{tabular}
\caption{Cross sections for the different models and for the dominant decay mode for $m_{H^\pm} = m_A = m_H = M = 300$  GeV and four values of $\tan \beta$. Remaining parameters are $m_h = 120$  GeV and $\sin (\beta - \alpha) = 1$.}
\end{center}
\vskip -0.6cm
\end{table}
\label{tab:B2}
In table 1 we present cross section values before (second column) and after (third to sixth column) the branching fraction for each mode are taken into account. A study based on $H^+ \to \tau \nu$ was performed by both ATLAS and CMS~\cite{ATLASCMS}. Using the ATLAS results for $m_{H^\pm} = 100$ GeV, the values of $\tan \beta = 1, 3$ and $7$ will be probed at 95 \% CL with less than 10 fb$^{-1}$ at $\sqrt{s} = 14$  TeV. Contrary to the MSSM (and models II and Y),  very large values of $\tan \beta$ cannot be probed with this process because the cross section in these models falls as $1/ \tan^2 \beta$. We now move to the mass benchmark of 300 GeV where all models are allowed. In tables 2 and 3 we present the cross sections for the charged Higgs main decay mode in the SM-like limit, that is, $\sin (\beta - \alpha) = 1$. It is clear from the tables that searches based on $H^+ \to \tau \nu$ cannot be used in this case. In fact, using the ATLAS study, only in the mass point $m_{H^\pm}=175$  GeV are we allowed to probe values of $\tan \beta$ of $\approx$ 4 and only for model X at 95 \% CL with 30 fb$^{-1}$ of collected luminosity at $\sqrt{s} = 14$ TeV (the MSSM and model II can be probed for very large values of $\tan \beta$). Furthermore, it is also clear that the models can only be distinguished for very large values of $\tan \beta$ and that there are some slim chances of distinguishing between models II and Y. A comparison of the signatures in the MSSM and in Model II was performed in \cite{Kanemura:2009mk}. 
\begin{table}[ht]
\begin{center}
\begin{tabular}{c c c c c c c c c c c c c c c c c c c c c c c c } \hline \hline
$\tan \beta$ && $\sigma_{I,X}$ (fb) && $\sigma_{II,Y}$ (fb)   &&  I (fb) &&  II  (fb)   && X (fb)  &&  Y  (fb)   \\
\hline
1        &&  2189  &&  2178    &&    1576 ($tb$)     &&  1568 ($tb$)        &&   1576 ($tb$)        &&  1568  ($tb$)       \\ \hline
3        &&  244    &&  247      &&    190 ($HW$)      &&  190 ($HW$)        &&   190 ($HW$)         && 190 ($HW$)       \\ \hline
7        &&  44      &&  72        &&    42 ($HW$)        &&  66 ($HW$)          &&   41 ($HW$)           && 66 ($HW$)      \\ \hline
30      &&  2.4     &&  562      &&    2.4 ($HW$)       &&  309 ($HW$)        &&   2.0 ($HW$)          && 348 ($HW$)      \\ \hline \hline
\end{tabular}
\caption{Cross sections for the different models and for the dominant decay mode for $m_{H^\pm} = m_A = 300$  GeV and four values of $\tan \beta$. Remaining parameters are $m_H = M = 130$ GeV, $m_h = 120$ GeV and $\sin (\beta - \alpha) = 1$.}
\end{center}
\vskip -0.6cm
\end{table}
\label{tab:B3}
Finally, in table 4 we present a scenario where $\sin (\beta - \alpha) = 0$ which allows a wider $\tan \beta$ range. Here, as in the remaining tables for 300 GeV, the Higgs decays almost exclusively to $tb$ for $\tan \beta = 1$. For larger values of $\tan \beta$ the sum of the branching ratios to $W S$, with $S=h,H,A$, is almost always close to 100 \%. As the neutral Higgs decays are also model dependent, a dedicated study for each case is needed (\cite{us}).  
\begin{table}[ht]
\begin{center}
\begin{tabular}{c c c c c c c c c c c c c c c c c c c c c c c c } \hline \hline
$\tan \beta$ && $\sigma_{I,X}$ (fb) && $\sigma_{II,Y}$ (fb)   &&  I (fb) &&  II  (fb)   && X (fb)  &&  Y  (fb)   \\
\hline
1        &&  2189  &&  2178    &&    1394 ($tb$)     &&  1387 ($tb$)     &&   1394 ($tb$)   &&  1387 ($tb$)    \\ \hline
3        &&  244    &&  247      &&    204 ($AW$)       &&  206 ($AW$)       &&   204 ($AW$)     && 206 ($AW$)       \\ \hline
7        &&  44      &&  72        &&    42 ($AW$)        &&  68 ($AW$)         &&   42 ($AW$)       && 66 ($AW$)      \\ \hline
30      &&  2.4     &&  562      &&    2.4 ($AW$)        &&  363 ($AW$)      &&   2.1 ($AW$)      && 393 ($AW$)      \\ \hline \hline
\end{tabular}
\caption{Cross sections for the different models and for the dominant decay mode for $m_{H^\pm} = m_H = 300$  GeV and four values of $\tan \beta$. Remaining parameters are $m_A = 100$  GeV, $m_h = 120$ GeV and $\sin (\beta - \alpha) = 0$. }
\end{center}
\vskip -0.6cm
\end{table}
\label{tab:B4}

\noindent
The most important contributions from the two next processes on our list \ref{Eq:bQH}, \ref{Eq:qQH} could only improve the limits obtained with the first process. The reason is that these contributions also depend only on the charged Higgs mass and on $\tan \beta$. Although our preliminary studies show that no major improvement can be achieved with processes \ref{Eq:bQH}, \ref{Eq:qQH}   a detailed study is still in progress~\cite{us}. A process that can clearly distinguish the MSSM for 2HDM, as shown in~\cite{WHb},  is the next process on the list $pp \to W^+ H^-$, but only when resonant production is allowed. In that case the main contribution to the cross section comes from the triangle diagram from the gluon fusion process.  The same is true for $pp \to H^+ H^-$ via gluon fusion, except that the region of the parameter space probed is complementary in the two processes~\cite{us}. Finally, a detailed study for the process $qQ   \to q'Q' H^+H^-$ in the 2HDM is in progress and will appear soon~\cite{us}.
\vskip -0.2cm

\section{Conclusion}

\noindent
We have presented a set of benchmarks for charged Higgs bosons searches in the 2HDM at the LHC. We have listed all production processes and discussed their importance in constraining the parameter space of the 2HDM and the different versions of the model. There are hints that in some scenarios, different versions of the 2HDM could be distiguished even with low luminosity with the LHC running at 14 TeV.


\begin{thebibliography}{99}

\bibitem{vacstab}
  P.~M.~Ferreira, R.~Santos and A.~Barroso,
  Phys.\ Lett.\  B {\bf 603} (2004) 219
  [Erratum-ibid.\  B {\bf 629} (2005) 114].

\bibitem{Glashow}  S.~L.~Glashow and S.~Weinberg,
  Phys.\ Rev.\  D {\bf 15}, 1958 (1977).

\bibitem{barger}
V.~D.~Barger, J.~L.~Hewett and R.~J.~N.~Phillips,
  Phys. \ Rev.\  D {\bf 41} (1990) 3421.

\bibitem{KY}
  M.~Aoki, S.~Kanemura, K.~Tsumura and K.~Yagyu,
  Phys.\ Rev.\  D {\bf 80} (2009) 015017.


\bibitem{Rhoparam}
A.~Denner, R.~J.~Guth, W.~Hollik and J.~H.~Kuhn,
  Z.\ Phys.\  C {\bf 51} (1991)  695.

\bibitem{pdg}
  K.~Nakamura {\it et al.}  [Particle Data Group],
  J.\ Phys.\ G {\bf 37} (2010) 075021.


\bibitem{osland} A.~Wahab El Kaffas, P.~Osland and O.~Magne Ogreid,
  Phys.\ Rev.\  D {\bf 76} (2007)  095001.


\bibitem{bsgamma}
%
  M.~Ciuchini, G.~Degrassi, P.~Gambino and G.~F.~Giudice,
  Nucl.\ Phys.\  B {\bf 527} (1998) 21
C.~Amsler \textit{et al.}, Phys. \ Lett. \ B {\bf 667}  (2008)  1, 
The Belle Collaboration, Phys. \ Rev. \ Lett. {\bf 103}  (2009) 241801.

\bibitem{LEP}
LEP Higgs Working Group for Higgs boson searches, ALEPH, DELPHI,L3 and OPAL Collaborations, arXiv: hep-ex/0107031.

\bibitem{Logan:2009uf}
 H.~E.~Logan and D.~MacLennan,
Phys.\ Rev.\  D {\bf 79} (2009) 115022.

\bibitem{vac1}
  N.~G.~Deshpande and E.~Ma,
  Phys.\ Rev.\  D {\bf 18} (1978) 2574.
  
 \bibitem{unitarity}
S.~Kanemura, T.~Kubota and E.~Takasugi,
Phys.\ Lett.\  B {\bf 313} (1993)  155; 
A.~G.~Akeroyd, A.~Arhrib and E.~M.~Naimi,
  Phys.\ Lett.\  B {\bf 490} (2000)  119.
 

\bibitem{btH} 
  J.~F.~Gunion, H.~E.~Haber, F.~E.~Paige, W.~K.~Tung and S.~S.~D.~Willenbrock,
  Nucl.\ Phys.\  B {\bf 294} (1987) 621;
 %
%
%
 J.~L.~Diaz-Cruz and O.~A.~Sampayo,
  Phys.\ Rev.\  D {50} (1994) 6820.


\bibitem{bQH}
  S.~Moretti and K.~Odagiri,
  Phys.\ Rev.\  D {55} (1997) 5627.


 
\bibitem{cs} 
        H.~J.~He and C.~P.~Yuan,
        Phys.\ Rev.\ Lett.\  {\bf 83} (1999) 28;
  S.~R.~Slabospitsky,
  arXiv:hep-ph/0203094;
  S.~Dittmaier, G.~Hiller, T.~Plehn and M.~Spannowsky,
  Phys.\ Rev.\  D {\bf 77} (2008) 115001.




\bibitem{WH}
  D.~A.~Dicus, J.~L.~Hewett, C.~Kao and T.~G.~Rizzo,
  Phys.\ Rev.\  D {40} (1989) 787;
%
  A.~A.~Barrientos Bendezu and B.~A.~Kniehl,
  Phys.\ Rev.\  D {59} (1999) 015009;
%
  S.~Moretti and K.~Odagiri,
  Phys.\ Rev.\  D {59} (1999) 055008;
%
  A.~A.~Barrientos Bendezu and B.~A.~Kniehl,
  Phys.\ Rev.\  D {61} (2000) 097701;
%
  O.~Brein, W.~Hollik and S.~Kanemura,
  Phys.\ Rev.\  D {\bf 63} (2001) 095001;
%
  D.~Eriksson, S.~Hesselbach and J.~Rathsman,
  Eur.\ Phys.\ J.\  C {\bf 53} (2008) 267;
%
  M.~Hashemi,
  arXiv:1008.3785 [hep-ph].

\bibitem{WHb}
  E.~Asakawa, O.~Brein and S.~Kanemura,
  Phys.\ Rev.\  D {72} (2005) 055017.
   
%




\bibitem{HpHm}
  S.~S.~D.~Willenbrock,
  Phys.\ Rev.\  D {35} (1987) 173;
%
  Y.~Jiang, L.~Han, W.~G.~Ma, Z.~H.~Yu and M.~Han,
  J.\ Phys.\ G {23} (1997) 385;
%
  A.~Krause, T.~Plehn, M.~Spira and P.~M.~Zerwas,
  Nucl.\ Phys.\  B {519} (1998) 85;
%
  A.~A.~Barrientos Bendezu and B.~A.~Kniehl,
  Nucl.\ Phys.\  B {568} (2000) 305;
%
  O.~Brein and W.~Hollik,
  Eur.\ Phys.\ J.\  C {13} (2000) 175;
%
  A.~Alves and T.~Plehn,
  Phys.\ Rev.\  D {\bf 71} (2005) 115014.






\bibitem{DY}
  E.~Eichten, I.~Hinchliffe, K.~D.~Lane and C.~Quigg,
  Rev.\ Mod.\ Phys.\  {56} (1984) 579
  [Addendum-ibid.\  {58} (1986) 1065].

\bibitem{VBF}
  S.~Moretti,
  J.\ Phys.\ G {28} (2002) 2567.



\bibitem{ATLASCMS}
  G.~Aad {\it et al.}  [The ATLAS Collaboration],
  arXiv:0901.0512 [hep-ex]; 
CMS Collaboration, Physics Technical Design Report, Volume 2. CERN/LHCC 2006- 021.

\bibitem{Kanemura:2009mk}
  S.~Kanemura, S.~Moretti, Y.~Mukai, R.~Santos and K.~Yagyu,
  Phys.\ Rev.\  D {\bf 79} (2009) 055017.

\bibitem{us}
 M.~Aoki, R.~Guedes, S.~Kanemura, S.~Moretti, R.~Santos and K.~Yagyu,
in preparation.

\end{thebibliography}
\end{document}